\documentclass[11pt]{article}  
\usepackage{latexsym}
\usepackage{graphicx}

\def\re{{\rm Re}}
\def\im{{\rm Im}}

\def\[{\left [}
\def\]{\right ]}
\def\({\left (}
\def\){\right )}
\def\bl{\bar{\lambda}}

\def\pp{\partial}

\def\G{{\cal G}}

\newcommand{\pr}{{\it Phys.\ Rev. }}
\newcommand{\pl}{{\it Phys.\ Lett. }}
\newcommand{\np}{{\it Nucl.\ Phys. }}

\def\L{{\cal L}}

\def\Tev{{\rm TeV}}
\def\Gev{{\rm GeV}}
\newcommand{\be}{\begin{equation}}
\newcommand{\ee}{\end{equation}}
\newcommand{\bea}{\begin{eqnarray}}
\newcommand{\eea}{\end{eqnarray}}

\begin{document}
\begin{titlepage}

      \hfill  LBNL-48640 

      \hfill  UCB-PTH-01/26 

      \hfill hep-ph/0108022

\hfill July 2001 \\[.2in]

\begin{center}

{\large \bf Progress in weakly coupled string phenomenology}\footnote{Talk 
presented at Spacetime Odyssey 2001, University of 
Michigan, Ann Arbor, MI, May 21-14, 2001, to be published in the 
proceedings.}\footnote{This work was 
supported in part by the Director, Office of
Energy Research, Office of High Energy and Nuclear Physics, Division
of High Energy Physics of the U.S. Department of Energy under
Contract DE-AC03-76SF00098 and in part by the National Science
Foundation under grants PHY-95-14797 and INT-9910077.}

Mary K. Gaillard \\[.05in]

{\em Department of Physics,University of California, and
 Theoretical Physics Group, 50A-5101, Lawrence Berkeley National Laboratory,
   Berkeley, CA 94720, USA}\\[.2in] 
\end{center}

\begin{abstract} The weakly coupled vacuum of $E_8\otimes E_8$ heterotic
string theory remains an attractive scenario for particle physics.
The particle spectrum and the issue of dilaton stabilization are
reviewed.  A specific model for hidden sector condensation and
supersymmetry breaking, that respects known constraints from string
theory, is described, and its phenomenological and cosmological
implications are discussed.

\end{abstract}
\end{titlepage}

\newpage
\renewcommand{\thepage}{\roman{page}}
\setcounter{page}{2}
\mbox{ }

\vskip 1in

\begin{center}
{\bf Disclaimer}
\end{center}

\vskip .2in

\begin{scriptsize}
\begin{quotation}
This document was prepared as an account of work sponsored by the United
States Government.  While this document is believed to contain
correct information, neither the United States Government nor any agency
thereof, nor The Regents of the University of California, nor any of their
employees, makes any warranty, express or implied, or assumes any legal
liability or responsibility for the accuracy, completeness, or usefulness
of any information, apparatus, product, or process disclosed, or represents
that its use would not infringe privately owned rights.  Reference herein
to any specific commercial products process, or service by its trade name,
trademark, manufacturer, or otherwise, does not necessarily constitute or
imply its endorsement, recommendation, or favoring by the United States
Government or any agency thereof, or The Regents of the University of
California.  The views and opinions of authors expressed herein do not
necessarily state or reflect those of the United States Government or any
agency thereof, or The Regents of the University of California.
\end{quotation}
\vfill

\end{scriptsize}

\vskip 2in

\begin{center}
\begin{small}
{\it Lawrence Berkeley Laboratory is an equal opportunity employer.}
\end{small}
\end{center}

\newpage
\renewcommand{\thepage}{\arabic{page}}
\setcounter{page}{1}

\section{Introduction: Approaches to the Theory of Everything}
\setcounter{equation}{0}
These days many theorists like to think that they are, in one way or another,
working on the Theory of Everything (ToE). There are two basic approaches.

\subsection{ Bottom Up}
The first approach starts with experimental data with the aim of deciphering
what it implies for an underlying, more fundamental theory. Its practitioners
are usually called phenomenologists. One outstanding
datum is the observed large gauge hierarchy, {\it i.e.,} the ratio of the $Z$ mass,
characteristic of the scale of electroweak symmetry breaking, to the reduced
Planck scale $m_P$:
$$ m_Z\approx 90\Gev \ll m_P = \sqrt{8\pi\over
G_N}\approx2\times10^{18}\Gev, $$ which can be technically resolved by
supersymmetry (SUSY) (among other conjectures--that have been tightly
constrained by experimental data, close to the point of being excluded
as relevant to the gauge hierarchy).  In addition, the conjunction of
SUSY and general relativity (GR) inexorably implies supergravity
(SUGRA).  The absence of observed SUSY partners (sparticles) requires broken SUSY
in the vacuum, and a more detailed analysis of the observed particle
spectrum constrains the mechanism of SUSY-breaking in the observable
sector: spontaneous SUSY-breaking is not viable, leaving soft
SUSY-breaking as the only option that preserves the technical SUSY
solution to the hierarchy problem.  This means introducing
SUSY-breaking operators of dimension three or less--such as gauge
invariant masses--into the Lagrangian for the SUSY extension of the
Standard Model (SM).  The unattractiveness of these {\it ad hoc} soft
terms strongly suggests that they arise from spontaneous SUSY breaking
in a ``hidden sector'' of the underlying theory.  Based on the above
facts, a number of standard scenarios have emerged.  These include:
\newline
$\bullet$ Gravity mediated SUSY-breaking, usually understood as
``Minimal SUGRA'' (MSUGRA), with masses of fixed spin particles set
equal at the Planck scale; this scenario is typically characterized
by\newline
$ m_{\rm scalars}= m_{0}> m_{\rm gauginos} = m_{1\over2}
\sim m_{\rm gravitino} = m_{3\over2}$ at the weak scale.
\newline
$\bullet$ Anomaly mediated SUSY-breaking~\cite{rs,hit}, in which $m_0
= m_{1\over2}=0$ classically; these models are characterized by $
m_{3\over2} >> m_0 ,\; m_{1\over2},$ and typically $m_0 >
m_{1\over2}$.  An exception is the Randall-Sundrum (RS) ``separable
potential'', constructed~\cite{rs} to mimic SUSY-breaking on a brane
spatially separated from our own in a fifth dimension; in this
scenario $m_0^2 < 0$ and $m_0$ arises first at two loops. More generally,
the scalar masses at one loop depend on the details of Planck-scale
physics~\cite{bgn}.
\newline
$\bullet$ Gauge mediated SUSY uses a hidden sector that has
renormalizable gauge interactions with the SM particles.  These
scenarios are typically characterized by small $m_{1\over2}$.
\begin{figure}[b]
\begin{picture}(350,220)(0,20)
\put(0,36){D = 9}\put(0,116){D = 10}\put(0,196){D = 11}
\multiput(120,40)(160,0){2}{\circle{40}}
\multiput(80,120)(80,0){4}{\circle{40}} \put(200,200){\circle{40}}
\put(68,104){\line(5,-6){40}}\put(260,120){\vector(1,-3){20}}
\put(107,57){\vector(1,-1){1}}\put(293,57){\vector(-1,-1){1}}
\put(332,104){\line(-5,-6){40}}\put(140,120){\vector(-1,-3){20}}
\put(184,188){\vector(-1,-2){24}}\put(216,188){\vector(1,-2){24}}
\put(188,184){\line(0,-1){80}}\put(212,184){\line(0,-1){80}}
\put(188,104){\vector(-1,-1){52}}\put(212,104){\vector(1,-1){52}}
\put(200,40){\vector(-1,0){60}}\put(200,40){\vector(1,0){60}}
\put(262,120){\vector(-1,0){2}}\put(262,136){\vector(-1,0){10}}
\put(304,134){\vector(0,-1){2}}\put(336,134){\vector(0,-1){2}}
\put(262,128){\oval(16,16)[r]}\put(320,134){\oval(32,32)[t]}
\put(222,116){WCHS}\put(304,116){$O(32)$}
\put(298,134){I}\put(340,134){H}\put(194,196){M}
\put(116,36){II}\put(270,36){H/I}\put(72,116){IIB}\put(152,116){IIA}
\put(88,80){\circle{8}}\put(130,90){\circle{8}}\put(220,164){\line(1,0){16}}
\put(312,80){\circle{8}}\put(270,90){\circle{8}}\put(172,164){\circle{8}}
\put(162,78){\oval(8,4)}\put(162,78){\oval(16,8)}
\put(230,78){\oval(4,8)}\put(246,78){\oval(4,8)[r]}
\put(230,82){\line(1,0){16}}\put(230,74){\line(1,0){16}}
\put(180,20){$T\leftrightarrow 1/T$}\put(242,140){$T\leftrightarrow
1/T$} \put(300,156){$S\leftrightarrow 1/S$}
\end{picture}
\caption{M-theory according to John Schwarz.
\label{fig:john}}
\end{figure}
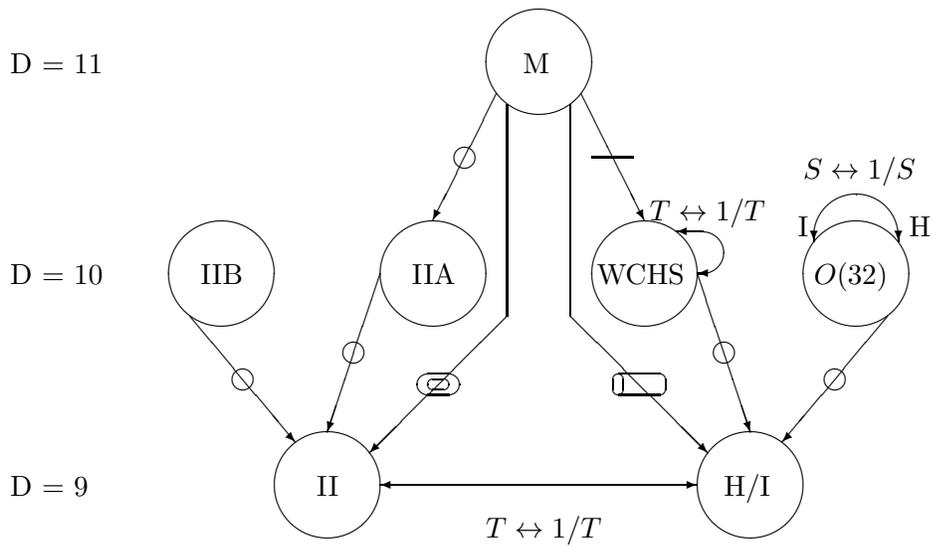
\begin{figure}[t]
\begin{picture}(350,215)(0,40)
\put(150,205){IIA,IIB: D-Branes}\put(120,200){\line(1,0){160}}
\multiput(160,190)(40,0){2}{$\times$}
\put(120,180){\oval(60,40)[l]}\put(280,190){\oval(40,20)[r]}
\put(120,140){\oval(40,40)[r]}\put(280,155){\oval(70,50)[l]}
\put(120,120){\line(-1,0){10}}\put(110,105){\oval(60,30)[l]}
\put(110,80){\oval(30,20)[r]}
\put(110,60){\oval(30,20)[l]}\put(280,110){\oval(80,40)[r]}
\put(280,90){\line(-1,0){10}}\put(270,75){\oval(40,30)[l]}
\put(270,55){\oval(40,10)[r]}\put(110,50){\line(1,0){160}}
\put(230,155){$\times$}\put(255,155){$O(32)_{\rm I}$}
\put(305,110){$\times$}\put(330,110){$O(32)_{\rm H}$}
\put(90,105){$\times$}\put(0,105){11-D SUGRA}\put(190,120){M}
\put(275,53){$\times$}\put(260,40){$E_8\otimes E_8$ WCHS}
\put(105,55){$\times$}\put(0,40){HW theory: {\small (very?) 
large extra dimension(s)}}\end{picture}
\caption{M-theory according to Mike Green.
\label{fig:mike}}\end{figure}
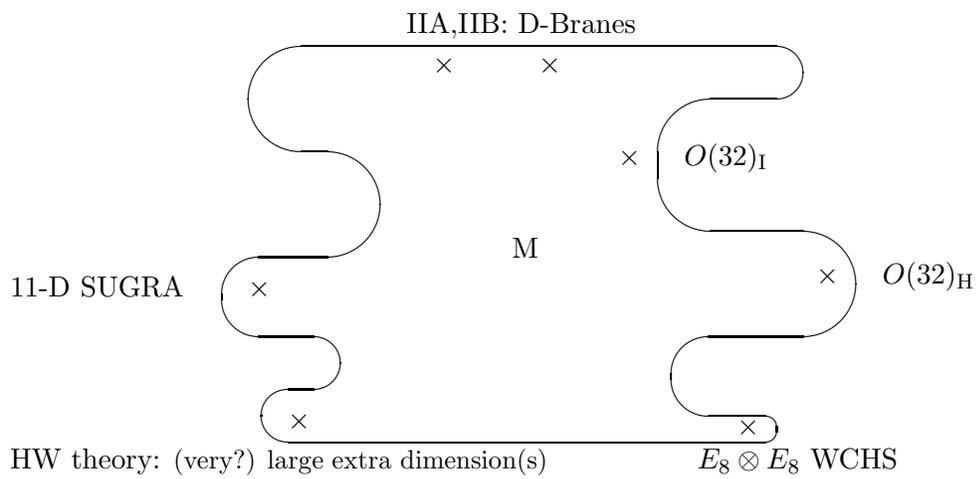

\subsection{Top Down} 
This approach starts from the ToE with the hope of deriving the
Standard Model from it; these days most of its practitioners are known
as string theorists.  The driving motivation is that superstring
theory is at present the only known candidate for reconciling GR with
quantum mechanics. These theories are consistent in ten dimensions;
over the last several years it was discovered that all the consistent
superstring theories are related to one another by dualities.  These
are, in my nomenclature: S-duality: $\alpha\to 1/\alpha,$ and
T-duality: Radius $\to 1/{\rm Radius},$ where $\alpha$ is the fine
structure constant of the gauge group(s) at the string scale, and
``Radius'' is a radius of compactification from dimension D to
dimension ${\rm D} -1$.  Figure~\ref{fig:john} shows~\cite{john} how
these dualities relate the various 10-D superstring theories to one
another, and to the currently presumed ToE, M-theory.  Not a lot else
is known about M-theory, except that it lives in 11 dimensions and
involves membranes.  In Figure~\ref{fig:john} the small circles, line,
torus and cylinder represent the relevant compact manifolds in
reducing D by one or two.  The two $O(32)$ theories are S-dual to one
another, while the $E_8\otimes E_8$ weakly coupled heterotic string
theory (WCHS) is perturbatively invariant~\cite{mod} under T-duality.
We will be specifically concentrating on this theory, and T-duality
will play an important role.

Another image of M-theory, shown in Figure~\ref{fig:mike}, which I call the
``puddle diagram'', indicates~\cite{mike} that all the known
superstring theories, as well as \mbox{D $=11$} SUGRA, are particular limits
of M-theory.  Currently, there is a lot of activity in type I and II
theories, or more generally in theories with branes.  Similarly the
Ho\v rava-Witten (HW) scenario~\cite{hw} and its inspirations have
received considerable attention.  If one compactifies one dimension of
the 11-D limit of M-theory, one gets the HW scenario with two 10-D
branes, each having an $E_8$ gauge group.  As the radius of this 11th
dimension is shrunk to zero, the WCHS scenario is recovered.  This is
the scenario addressed here, in a marriage of the two approaches that
may serve as an illustrative example of the diversity of
possible SUSY breaking scenarios.

\section{The $E_8\otimes E_8$ Heterotic String}\setcounter{equation}{0}
Let us recall the reasons for the original appeal of the weakly coupled
$E_8\otimes E_8$ heterotic string theory~\cite{gross} compactified on
a Calabi-Yau (CY) manifold~\cite{cy} (or a CY-like
orbifold~\cite{orb}).  The zero-slope (infinite string tension) limit
of superstring theory~\cite{gs} is ten dimensional supergravity
coupled to a supersymmetric Yang-Mills theory with an $E_8\otimes E_8$
gauge group.  To make contact with the real world, six of these ten
dimensions must be compact--of size much smaller than distance
scales probed by particle accelerators, and generally assumed to be of
order of the reduced Planck length, $10^{-32}$cm.  If the topology of
the extra dimensions were a six-torus, which has a flat geometry, the
8-component spinorial parameter of $N=1$ supergravity in ten
dimensions would appear as the four two-component parameters of $N=4$
supergravity in ten dimensions.  However a Calabi-Yau
manifold leaves only one of these spinors invariant under parallel
transport; for this manifold the group of transformations under
parallel transport (holonomy group) is the $SU(3)$ subgroup of the
maximal $SU(4) \cong SO(6)$ holonomy group of a six dimensional
compact space.  This breaks $N=4$ supersymmetry to $N=1$ in four
dimensions.  As is well known, the only phenomenologically viable
supersymmetric theory at low energies is $N=1$, because it is the only
one that admits complex representations of the gauge group that are
needed to describe quarks and leptons. For this solution, the
classical equations of motion impose the identification of the affine
connection of general coordinate transformations on the compact space
(described by three complex dimensions) with the gauge connection of
an $SU(3)$ subgroup of one of the $E_8$'s: $E_8\ni E_6\otimes SU(3)$,
resulting in $E_6\otimes E_8$ as the gauge group in four dimensions. 
Since the early 1980's, $E_6$ has been considered the largest
group that is a phenomenologically viable candidate for a Grand
Unified Theory (GUT) of the Standard Model.  Hence $E_6$ is
identified as the gauge group of the ``observable sector'', and the
additional $E_8$ is attributed to a ``hidden sector'', that interacts
with the former only with gravitational strength couplings.
Orbifolds, which are flat spaces except for points of infinite curvature, are
more easily studied than CY manifolds, and orbifold compactifications that
closely mimic the CY compactification described above, and that yield
realistic spectra with just three generations of quarks and leptons, have been
found~\cite{iban}.  In this case the surviving gauge group is $E_6\otimes\G_o
\otimes E_8,\;\G_o\in SU(3)$. 
The low energy effective field theory is determined by the massless
spectrum, {\it i.e.}, the spectrum of states with masses very small
compared with the scales of the string tension and of
compactification. Massless bosons have zero triality under an $SU(3)$
which is the diagonal of the $SU(3)$ holonomy group and the (broken)
$SU(3)$ subgroup of one $E_8$.  The ten-dimensional vector fields
$A_M,\; M = 0,1,\ldots 9,$ appear in four dimensions as four-vectors
$A_\mu,\;\mu = M = 0,1,\ldots 3$, and as scalars $A_m,\; m = M-3 =
1,\cdots 6.$ Under the decomposition $E_8\ni E_6\otimes SU(3)$, the
$E_8$ adjoint contains the adjoints of $E_6$ and $SU(3)$, and the
representation ${\bf(27,3)} + {\bf(\overline{27},\overline{3})}$.
Thus the massless spectrum includes gauge fields in the adjoint
representation of $E_6\otimes\G_o\otimes E_8$ with zero triality under
both $SU(3)$'s, and scalar fields in ${\bf 27 + \overline{27}}$ of
$E_6$, with triality $\pm1$ under both $SU(3)$'s, together with their
fermionic superpartners.  The number of ${\bf 27}$ and
${\bf\overline{27}}$ chiral supermultiplets that are massless depends
on the detailed topology of the compact manifold.  The important point
for phenomenology is the decomposition under $E_6\to SO(10)\to SU(5)$:
\be ({\bf 27})_{E_6} = ({\bf 16 + 10 + 1})_{SO(10)} = \({\bf \{\bar{5}
+ 10 + 1\} + \{5 + \bar{5}\} + 1}\)_{SU(5)}.\ee 
A ${\bf \overline{5} +
10 + 1}$ contains one generation of quarks and leptons of the Standard
Model, a right-handed neutrino and their scalar superpartners; a ${\bf
5 + \overline{5}}$ contains the two Higgs doublets needed in the
supersymmetric extension of the Standard Model and their fermion
superpartners, as well as color-triplet supermultiplets. Thus all the
states of the Standard Model and its minimal supersymmetric
extension are present.
On the other hand, there are no scalar particles in the adjoint
representation of the gauge group. In conventional models for grand
unification, these (or one or more other representations much larger
than the fundamental one) are needed to break the GUT group to the
Standard Model.  In string theory, this symmetry breaking can be
achieved by the Hosotani, or ``Wilson line'', mechanism~\cite{hos} in
which gauge flux is trapped around ``holes'' or ``tubes'' in the
compact manifold, in a manner reminiscent of the Arahonov-Bohm effect.
The vacuum value of the trapped flux $<\int d\ell^m A_m>$ has the same
effect as an adjoint Higgs, without the complications of having to
construct a potential for large Higgs representations that can
actually reproduce the properties of the observed vacuum~\cite{lang}.
When this effect is included, the gauge group in four dimensions is
\bea &&\G_{obs}\otimes\G_{hid},
\quad\G_{obs}=\G_{SM}\otimes\G'\otimes\G_o,\quad \G_{SM}\otimes\G'\in
E_6, \quad \G_o\in SU(3),\nonumber \\ && \G_{hid}\in E_8,\quad \G_{SM}
= SU(3)_c\otimes SU(2)_L\otimes U(1)_w.
\label{eq:group}\eea

There are many other four dimensional string vacua in addition to the
class of vacua described above. However the attractiveness of that
picture is that the requirement of $N=1$ SUSY naturally results in a
phenomenologically viable gauge group and particle spectrum. Moreover,
the gauge symmetry can be broken to a product group embedding the
Standard Model without the necessity of introducing large Higgs
representations.  In addition, the $E_8\otimes E_8$ string theory
provides a hidden sector needed for a viable theory of spontaneously
broken SUSY.  More specifically, if some subgroup $\G_a$ of $\G_{hid}$
is asymptotically free, with a $\beta$-function coefficient
$b_a>b_{SU(3)}$, defined by the renormalization group equation (RGE)
\be \mu{\pp g_a(\mu)\over\pp\mu} = -{3\over2}b_ag_a^3(\mu) +
O(g_a^5)\label{eq:rge},\ee confinement and fermion condensation will
occur at a scale $\Lambda_c\gg\Lambda_{QCD}$, and hidden sector
gaugino condensation $<\bl\lambda>_{\G_a} \ne 0,$ may
induce~\cite{nilles} supersymmetry breaking.  To discuss supersymmetry
breaking in more detail, we need the low energy spectrum resulting
from the ten-dimensional gravity supermultiplet that consists of the
10-D metric $g_{MN}$, an antisymmetric tensor $b_{MN}$, the dilaton
$\phi$, the gravitino $\psi_M$ and the dilatino $\chi$.  For the class
of CY and orbifold compactifications described above, the massless
bosons in four dimensions are the 4-D metric $g_{\mu\nu}$, the
antisymmetric tensor $b_{\mu\nu}$, the dilaton $\phi$, and certain
components of the tensors $g_{mn}$ and $b_{mn}$ that form the real and
imaginary parts, respectively, of complex scalars known as moduli.
(More precisely, the scalar components of the chiral multiplets of the
low energy theory are obtained as functions of the scalars
$\phi,g_{mn}$, while the pseudoscalars $b_{mn}$ form axionic
components of these supermultiplets.)  The number of moduli is related
to the number of particle generations (\# of ${\bf 27}$'s $-$ \# of
${\bf\overline{27}}$'s).  Typically, in a three generation orbifold
model there are three moduli $t_I$; the $vev$'s $<{\rm Re}t_I>$
determine the radii of compactification of the three tori of the
compact space.  In some compactifications there are three other moduli
$u_I$; the $vev$'s $<{\rm Re}u_I>$ determine the ratios of the two
{\it a priori} independent radii of each torus.  These form chiral
multiplets with fermions $\chi^t_I, \chi^u_I$ obtained from components
of $\psi_m$.  The 4-D dilatino $\chi$ forms a chiral multiplet with
with a complex scalar field $s$ whose $vev$ $ <s> = g^{-2} -
i\theta/8\pi^2$ determines the gauge coupling constant and the
$\theta$ parameter of the 4-D Yang-Mills theory.  The ``universal''
axion Im$s$ is obtained by a duality transformation~\cite{wit} from the
antisymmetric tensor $b_{\mu\nu}$:  $\pp_\mu{\rm Im}s\leftrightarrow
\epsilon_{\mu\nu\rho\sigma}\pp^\nu b^{\rho\sigma}.$ Because the
dilaton couples to the (observable and hidden) Yang-Mills sector,
gaugino condensation induces~\cite{dine} a superpotential for the
dilaton superfield\footnote{Throughout I use capital Greek or Roman
letters to denote a chiral superfield, and the corresponding lower
case letter to denote its scalar component.} $S$: \be W(S)\propto
e^{-S/b_a}.\label{eq:dil}\ee The vacuum value $ <W(S)> \propto
\left<e^{-S/b_a}\right> = e^{-g^{-2}/b_a}= \Lambda_c$ is governed by
the condensation scale $\Lambda_c$ as determined by the RGE
(\ref{eq:rge}).  If it is nonzero, the gravitino acquires a mass
$m_{3\over2}\propto<W>$, and local supersymmetry is broken.

\section{The Runaway Dilaton: A Brief Abridged
History}\setcounter{equation}{0}

The superpotential (\ref{eq:dil}) results in a potential for the dilaton of the
form $ V(s)\propto e^{-2\re s/b_a},$ which has its minimum at
vanishing vacuum energy and vanishing gauge coupling: $<\re s>
\to\infty,\; g^2\to 0$.  This is the notorious runaway dilaton
problem.  The effective potential for $s$ is in fact determined from anomaly
matching~\cite{vy}: $\delta\L_{eff}(s,u) \longleftrightarrow
\delta\L_{hid}({\rm gauge}),$ where $u, \;\left<u\right> = \left<
\bl\lambda\right>_{\G_a},$ is the lightest scalar bound state of the
strongly interacting, confined gauge sector.  Just as in QCD, the
effective low energy theory of bound states must reflect both the
symmetries and the anomalies--quantum induced breaking of classical
symmetries--of the underlying Yang-Mills theory.  It turns out that
the effective quantum field theory (QFT) is anomalous under
T-duality.  Since this is an exact symmetry of
heterotic string perturbation theory, it means that the effective QFT
is incomplete.  This is cured by including model dependent string-loop
threshold corrections~\cite{thresh} as well as a ``Green-Schwarz''
(GS) counter-term~\cite{gsterm}, named in analogy to a similar anomaly
canceling mechanism in 10-D SUGRA~\cite{gs}.  This introduces
dilaton-moduli mixing, and the gauge coupling constant is now
identified as $g^2= 2\left<\ell\right>,\;\ell^{-1} =
2\re s - b{\sum_I} \ln(2\re t^I), $ 
where $b\le b_{E_8} = 30/8\pi^2 $ is the coefficient
of the GS term.  It also introduces a second runaway direction, this time
at strong coupling: $V \to - \infty$ for $g^2\to\infty$. The small
coupling behavior is unaffected, but the potential becomes
negative for $\alpha = \ell/2\pi > .57$. This is the strong coupling
regime, and nonperturbative string effects cannot be neglected; they
are expected~\cite{shenk} to modify the K\"ahler potential
for the dilaton, and therefore the potential $V(\ell,u)$.  It has been
shown~\cite{us,casas} that these contributions can indeed stabilize
the dilaton.

The remainder of this paper describes an explicit model~\cite{us}
based on affine level one\footnote{This is a simplifying but not a
necessary assumption.} orbifolds with three untwisted moduli $T_I$ and
a gauge group of the form (\ref{eq:group}).  Retaining just one or two
terms of the suggested parameterizations~\cite{shenk} of
the nonperturbative string corrections: $a_n\ell^{-n/2}e^{-c_n/\sqrt{\ell}}$
or $a_n\ell^{-n}e^{-c_n/\ell},$ the potential can be made
positive definite everywhere and the parameters can be chosen to fit
two data points: the coupling constant $g^2\approx 1/2$ and the
cosmological constant $\Lambda \simeq 0$.  This is fine tuning, but it
can be done with reasonable (order 1) values for the parameters $c_n,
a_n$.  If there are several condensates with different
$\beta$-functions, the potential is dominated by the condensate with
the largest $\beta$-function coefficient $b_+$, and the result is
essentially the same as in the single condensate case, except that a
small mass is generated for the axion Im$s$.

\section{Features of the Condensation Model}
\setcounter{equation}{0}
In this model, mass hierarchies arise from the presence of
$\beta$-function coefficients; these have interesting implications for
both cosmology and the spectrum of sparticles--the supersymmetric
partners of the SM particles.\footnote{The soft SUSY breaking
parameters were calculated in~\cite{us} for $<t_I> = 1$; the results
are similar if $<t_I> = e^{i\pi/6}$.}

\subsection{Modular Cosmology}

The masses of the dilaton $d=\re s$ and the
complex $t$-moduli are related to the gravitino mass by~\cite{us}
\be m_d \sim {1\over b_+^2} m_{\tilde G}, \quad m_{t^I} \approx
{2\pi\over3}{(b-b_+)\over(1+b<\ell>)}m_{\tilde G}.
\label{eq:modmass} \ee
Taking $b = b_{E_8} \approx .38 \approx 10b,$ gives a
hierarchy of order $m_{3\over2}\sim 10^{-15}m_{Pl}\sim 10^3GeV$ and
$m_{t_I}\approx 20 m_{3\over2}\approx20TeV,\;m_d\sim
10^3m_{2\over3}\sim 10^6GeV$, which is sufficient to evade the late
moduli decay problem~\cite{modprob} in nucleosynthesis.

If there is just one hidden sector condensate, the axion $a = \im s$
is massless up to QCD-induced effects:
$m_a\sim(\Lambda_{QCD}/\Lambda_c)^{3\over2}m_{3\over2}\sim10^{-9}eV$,
and it is the natural candidate for the Peccei-Quinn axion. Because of
string nonperturbative corrections to its gauge kinetic term, the
decay constant $f_a$ of the canonically normalized axion is reduced
with respect to the standard result by a factor $b_+ \ell^2
\sqrt{6}\approx 1/50$ if $b_+\approx .1b_{E_8}$, which may be
sufficiently small to satisfy the (looser) constraints on $f_a$ when
moduli are present~\cite{bd}.

\subsection{Sparticle Spectrum}

\begin{figure}[t]
\includegraphics{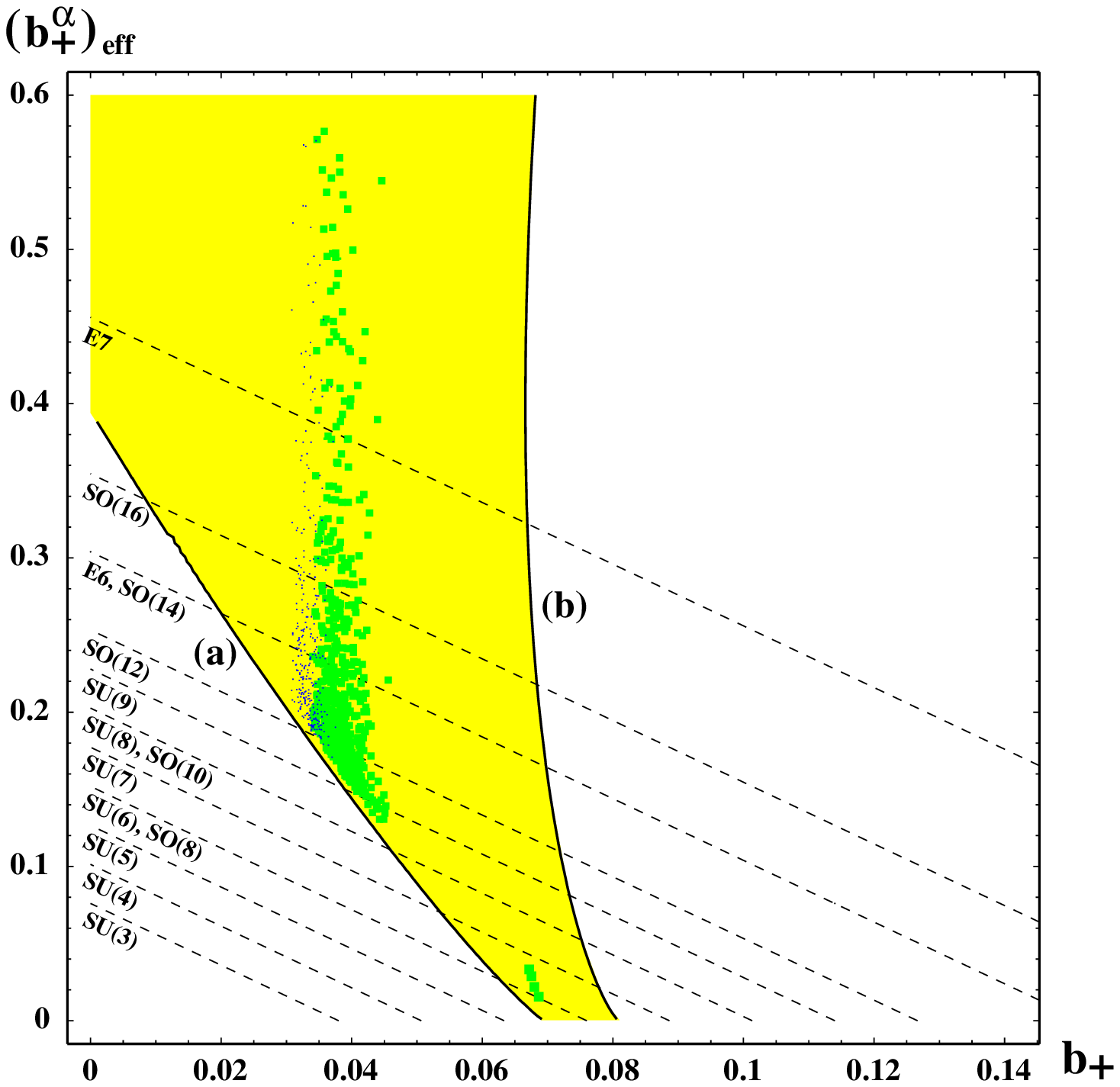}
\caption{Viable hidden sector gauge groups for
scenario A of the condensation model. The swath bounded
by lines (a) and (b) is the region defined by $.1<m_{3\over2}/\Tev,\lambda_c
<10$, where $\lambda_c$ is a condensate superpotential coupling constant.
The fine points correspond to $.1\le\Omega_dh^2\le.3$, and the
course points to $.3<\Omega_dh^2\le1$.
\label{fig:abn}}
\end{figure}

In contrast to an enhancement of the dilaton and moduli masses, there
is a suppression of gaugino masses: $m_{1\over2} \approx
b_+m_{3\over2}$, as evaluated at the scale $\Lambda_a$ in the tree
approximation.  As a consequence quantum corrections can be important;
for example there is an anomaly-like scenario in some regions of
the $(b_+,b_+^\alpha)$ parameter space, where $b_+^\alpha$ is the
hidden matter contribution to $b_+.$ If the gauge group for the
dominant condensate (largest $b_a$) is not $E_8$, the moduli $t_I$ are
stabilized through their couplings to twisted sector matter and/or
moduli-dependent string threshold corrections at a self-dual point,
and their auxiliary fields vanish in the vacuum. Thus SUSY-breaking is
dilaton mediated, avoiding a potentially dangerous source of flavor
changing neutral currents (FCNC).  These results hold up to unknown
couplings $p_A$ of chiral matter $\phi^A$ to the GS term: at the scale
$\Lambda_a$ $ m_{0A} = m_{3\over2}$ if $p_A = 0$, while $m_{0A}=
{1\over2}m_{t^I}\approx 10 m_{3\over2}$ if the scalars couple with the
same strength as the T-moduli: $ p_A = b$.  In addition, if $p_A = b$
for some gauge-charged chiral fields, there are enhanced loop
corrections to gaugino masses~\cite{gnw}.  Four sample scenarios were
studied~\cite{gn}: A) $p_A=0$, B) $p_A=b$, C) $p_A=0$ for the
superpartners of the first two generations of SM particles and $p_A=b$
for the third, and D) $p_A=0$ for the Higgs particles and $p_A=b$
otherwise.  Imposing constraints from experiments 
and the correct
electroweak symmetry-breaking vacuum rules out scenarios B and C.
Scenario A is viable for $1.65< \tan\beta<4.5$, and scenario D is
viable for all values of $\tan\beta$, which is the ratio of Higgs
$vev$'s in the supersymmetric extension of the SM that requires two
Higgs chiral multiplets.  The viable range of $(b_+,b_+^\alpha)$
parameter space is shown~\cite{abn} 
in Figure~\ref{fig:abn} for $g^2={1\over2}$.
The dashed lines represent the possible dominant condensing hidden
gauge groups $\G_+\in E_8$ with chiral matter in the coset space
$E_8/\G_{hid}$. 
\section{Other Issues in Cosmology}
\setcounter{equation}{0}
\subsection{Flat Directions in the Early Universe}

Many successful cosmological scenarios--such as an epoch of
inflation--require flat directions in the potential.  A promising
scenario for baryogenesis suggested~\cite{ad} by Affleck and Dine
(AD) requires in particular flat directions during inflation
in sparticle field space: $<\tilde q>, <\tilde\ell>\ne0$, where
$\tilde f$ denotes the superpartner of the fermion $f$.  While flat
directions are common in SUSY theories, they are generally
lifted~\cite{drt} in the early universe by SUGRA couplings to the
potential that drives inflation.  This problem is evaded~\cite{gmo} in
models with a ``no-scale'' structure, such as the classical potential
for the untwisted sector of orbifold compactifications. Although the
GS term breaks the no-scale property of the theory, quasi-flat
directions can still be found. An explicit model~\cite{lyth} for
inflation based on the effective theory described above allows
dilaton stabilization within its domain of
attraction with one or more moduli stabilized at the vacuum value
 $t_I=e^{i\pi/6}$. One of the moduli may be the inflaton.
The moduli masses (\ref{eq:modmass}) are sufficiently large to evade the
late moduli decay problem in nucleosynthesis, but unlike the dilaton,
they are insufficient to avoid a large relic LSP density without
violation~\cite{lsp} of R-parity (a quantum number that distinguishes
SM particles from their superpartners). If R-parity is conserved, this
problem can be evaded if the moduli are stabilized at or near their
vacuum values--or for a modulus that is itself the inflaton.  It is
possible that the requirement that the remaining moduli be in the
domain of attraction is sufficient to avoid the problem altogether.
For example, if $\im t_I = 0$, the domain of attraction near $t_I = 1$
is rather limited: $0.6<{\rm Re}t_I<1.6$, and the entropy produced by
dilaton decay with an initial value in this range might be less than
commonly assumed.  The dilaton decay to its true ground state may
provide~\cite{cgmo} partial baryon number dilution, which is generally
needed for a viable AD scenario.

\subsection{Relic Density of the Lightest SUSY Particle (LSP)}

Two pertinent questions for SUSY cosmology are:
\newline
$\bullet$ Does the LSP overclose the Universe?
\newline
$\bullet$ Can the LSP be dark matter?
\newline
As discussed by Joe Silk~\cite{Joe}, the window for LSP dark matter in
the much-studied MSUGRA scenario~\cite{efo}, has become ever more tiny
as the Higgs mass limit has increased; in fact there is not much
parameter space in which the LSP does not overclose the universe.  The
ratios of electroweak sparticle masses at the Plank scale determine the
composition of the LSP (which must be neutral) in terms of the Bino
(superpartner of the SM $U(1)$ gauge boson), the Wino (superpartner of
the SM $SU(2)$ gauge boson), and the higgsino (superpartner of the
Higgs boson).  The MSUGRA assumption of equal gaugino masses at the
Planck scale leads to a Bino LSP with rather weak couplings, resulting
in little annihilation and hence the tendency to overclose the
universe, except in a narrow range of parameter space where the LSP is
nearly degenerate with the next to lightest sparticle (in this case a
stau $\tilde\tau$), allowing significant coannihilation.  Relaxing
this assumption~\cite{abn} it was found that a predominantly Bino LSP
with a small admixture of Wino can provide the observed amount
$\Omega_d$ of dark matter.  In the condensation model, this occurs in
the region indicated by fine points in Figure~\ref{fig:abn}.  In this
model the deviation from the MSUGRA scenario is due to the importance
of loop corrections to small tree-level gaugino masses; in addition to
a small Wino component in the LSP, its near degeneracy in
mass with the lightest charged gaugino enhances
coannihilation.  For larger $b_+$ the LSP becomes pure Bino as in
MSUGRA, and for smaller values it becomes Wino-dominated as in
anomaly-mediated models which are cosmologically safe, but do not
provide LSP dark matter, because Wino annihilation is too fast.

\section{Issues:  Realistic Orbifold Models}
\setcounter{equation}{0}
Orbifold compactifications with
the Wilson line/Hosotani mechanism needed to break
$E_6$ to the SM gauge group generally have $b_+\le b\le b_{E_8}$. 
An example is a model~\cite{iban} with hidden gauge
group $O(10)$ and $b_+= b = b_{O(10)}$. It is clear from
(\ref{eq:modmass}) that this would lead to disastrous modular cosmology,
since the $t$-moduli are massless.  Moreover,
in typical orbifold compactifications, the gauge group
$\G_{obs}\otimes\G_{hid} $ obtained at the string scale has no
asymptotically free subgroup that could condense to trigger
SUSY-breaking. However in many compactifications with realistic
particle spectra~\cite{iban}, the effective field theory has an
anomalous $U(1)$ gauge subgroup, which is not anomalous at the string
theory level.  The anomaly is canceled~\cite{dsw} by a GS
counterterm, similar to the GS term introduced above to cancel the
modular anomaly.  This results in a D-term that forces some
otherwise flat direction in scalar field space to acquire a vacuum
expectation value, further breaking the gauge symmetry, and giving
masses of order $\Lambda_D$ to some chiral multiplets, so that the
$\beta$-function of some of the surviving gauge subgroups may be
negative below the scale $\Lambda_D$, typically an order of magnitude
below the string scale.  The presence of such a 
D-term was explicitly invoked in the above-mentioned inflationary
model~\cite{lyth}. Its incorporation into the effective condensation
potential is under study.

There is a large vacuum degeneracy associated with the D-term induced
breaking of the anomalous $U(1)$, resulting in many massless
``D-moduli'' that have the potential for a yet more disastrous modular
cosmology~\cite{joel}. However preliminary results indicate that the
D-moduli couplings to matter condensates lift the degeneracy to give
cosmologically safe D-moduli masses.  Although the D-term modifies the
potential for the dilaton, one still {\mbox obtains} moduli stabilized at
self-dual points giving FCNC-free dilaton dominated SUSY-breaking, an
enhanced dilaton mass $m_d$ and a suppressed axion coupling $f_d$.  An
enhancement of the ratio $m_{t^I}/m_{3\over2}$ can result from
couplings to condensates of $U(1)$-charged D-moduli, that also carry
T-modular weights.

\section{Conclusions}
\setcounter{equation}{0}
The lessons of this talk are three-fold:
\newline
$\bullet$ Quantitative studies with predictions for observable phenomena
are possible within the context of the WCHS.
\newline
$\bullet$ Experiments can place restrictions on the underlying theory,
such as the hidden gauge sector physics through restriction on the
allowed $(b_+,b_+^\alpha)$ parameter space, and the couplings and
modular weights of D-moduli when an anomalous $U(1)$ is present.
Experiments can also inform us about Plank scale physics, such as
matter couplings to the GS term.  The one-loop corrections to the soft
scalar potential are also sensitive to the details of Plank scale
physics.  
\newline
$\bullet$ Searches for sparticles should avoid restrictive assumptions!

\section*{Acknowledgments}
I am grateful to my many collaborators, and to
Joel Giedt and John Schwarz for discussions of string nonperturbative
effects.  This work was supported in part by the Director, Office of
Energy Research, Office of High Energy and Nuclear Physics, Division
of High Energy Physics of the U.S. Department of Energy under Contract
DE-AC03-76SF00098 and in part by the National Science Foundation under
grants PHY-95-14797 and INT-9910077.


\begin{thebibliography}{99}
\bibitem{mike} M. Green, a seminar; my apologies if the puddle diagram
was invented by someone else.
\end{thebibliography}
\end{document}